\documentclass[10pt,leqno]{amsart}
\usepackage{graphicx}
\usepackage{indentfirst,csquotes}

\usepackage{amssymb,amsthm,amsmath}
\usepackage{xcolor,paralist,hyperref,titlesec,fancyhdr,etoolbox}

\titleformat{\section}[display]{\normalfont\huge\bfseries\centering}{\centering\chaptertitlename\thechapter}{10pt}{\Large}
\titlespacing*{\section}{0pt}{0ex}{0ex}

\hypersetup{ colorlinks=true, linkcolor=black, filecolor=black, urlcolor=black }

\usepackage{lipsum}

\begin{document}
\title{Four-wave mixing mediated synchronization of localized polariton condensates} %%%%%%%%%%%%
\author[A.V. Yulin]{Alexey V. Yulin}
\date{\today}
\address{Department of Physics, ITMO University, Saint Petersburg 197101, Russia}
\email{alex.v.yulin@gmail.com}
\maketitle

\let\thefootnote\relax
% \footnotetext{MSC2020: Primary 00A05, Secondary 00A66.} %%%%%%%%%%

\begin{abstract}
The Letter is devoted to new frequencies generation and its role in the synchronization of two spatially separated polariton condensates affected by the coherent pump with the frequency detuned from the frequencies of the condensates. We focus on the case where the distance between the condensates is so long that their interaction through the evanescent tails is negligible. By numerical simulation we show that the four-wave mixing of the polaritons with the coherent drive produces the free propagating polaritons that can  mediate the inter-condensate interaction resulting in the phase-locking of the condensates.
\end{abstract} %%%%%%%%%

\bigskip

The physics of exciton polaritons has been attracting much attention since the pioneering works where polariton condensation is reported in optically \cite{kasprzak2006bose,balili2007bose} and electrically \cite{schneider2013electrically} pumped microcavities.  One of the areas of active research is spatially extended condensate, including the lattices of condensate droplets; for examples, see \cite{lai2007coherent,galbiati2012polariton,topfer2021engineering}. Speaking about practical importance of the polariton condensate it should be mentioned that apart from direct usage as tunable sources of coherent light the polaritons can find application in information procession, in particular it has been suggested to use polariton lattices for solving the classical spin model \cite{berloff2017realizing}. 

From the point of view of nonlinear dynamics, the formation of macroscopically coherent states of the polariton droplets is nothing else but synchronization of interacting nonlinear oscillators. The concept of synchronization has been developed in a number of works treating interacting polariton condensates sustained by an incoherent pump  \cite{wouters2008synchronized,baas2008synchronized,ohadi2018synchronization,topfer2020time,toebes2022dispersive}, including the case of periodically oscillating condensates \cite{christmann2014oscillatory,saito2016self,toebes2022dispersive,ramos2024theory}. The coherent pump can change synchronization dramatically, in particular due to the competition between the mutual locking of the phase of the condensates and their synchronization with the coherent pump \cite{kalinin2018simulating,chestnov2019optical}. 

In this Letter we consider the case where the polariton condensates can be phase locked neither by direct interaction though the evanescent tails no by phase locking to the common coherent pump. In the following, we discuss a new mechanism of synchronization based on the exchange of the polaritons generated via four-wave mixing of the condensates with the coherent pump. 

The Letter is structured as follows. First, we introduce a mathematical model describing the polariton condensate in terms of the generalized Gross-Pitaevskii equation. Then we consider a single condensate trapped in a potential well with some effective gain produced by the reservoir of incoherent excitons. We study the action of the frequency detuned coherent drive on the condensate and show that it can produce new frequencies via four-wave mixing with the condensate. Finally, we study two spatially separated condensates and show that the generated polaritons can establish coherence between the condensates. 

The main purpose of this work is a proof of concept and therefore we consider a simplest but physically relevant case where the polariton dynamics can be described by generalized Gross-Pitaevskii equation which we write in dimensionless units
\begin{eqnarray}
\frac{\partial A }{\partial t} =\frac{i}{2}\nabla^2 A+ \frac{1}{2} \left( \frac{1-2i\beta}{1+|A|^2}P -1 -iV -i\alpha |A|^2\right)  A + P_{ch} \label{main_eq}
\end{eqnarray}
where $A$ is a complex order parameter function describing the polariton condensate, $P(\vec r, t)$ is the intensity of the incoherent (optical or electric) pump supporting the polaritons, $V(\vec r, t)$ is the effective potential that can be controlled by the cavity design, $P_{ch}(\vec r, t)$ is coherent optical pump. The coefficient $\alpha$ accounts for nonlinear polariton-polariton interaction resulting in the polariton frequency blue shift, the coefficient $beta$ is the ratio of the polariton frequency blue-shift to the effective gain caused by the incoherent pump. 

Equation (\ref{main_eq}) can easily be derived from a more general model consisting of the equation for the polariton order parameter and the equation for the reservoir of incoherent excitons under the assumption that the exciton relaxation time is much shorted compared to the characteristic times of the polariton evolution. The time is normalized on the polariton life time $\tau$, the spatial coordinate on the characteristic length $L=\sqrt{ 2 D \tau}$ where $D$ is the dispersion of the polaritons, the incoherent pump is normalized on the condensation threshold. 
For the parameters taken from \cite{PhysRevB.94.134310} and the polariton dispersion $D=0.8$ meV$\cdot\mu$ m$ ^2$
we obtain $\tau=20$ ps, $L=7$ $\mu$m, $\alpha=4.56$, $\beta=1.82$ and allnumerical simulations discussed in the paper are done using these parameters.

First we consider four-wave mixing of the condensate with a coherent spatially uniform pump. For sake os simplicity we consider a one-dimensional problem and choose a trapping potential in a super-Gaussian form close to a rectangular potential well $V=V_0  \left( 1-\exp\left( -\frac{x^{12}}{w^{12}} \right)\right)$. We choose the potential depth $V_0=40$ and half width $0.15$ which provides that there is only one state localized in the potential. This choice insures that only polaritons in the ground state are excited for the other states experience high radiative losses. Let us remark that there are other means to ensure the creation of single frequency condensate states, however in this Letter we discuss only the simplest one. To achieve condensation of the polariutons we use a pump that we also take in the form of super-Gaussian function $P=\mu \exp\left( - \frac{x^{12}}{w_{\mu}^{12}} \right)$ where $\mu$ is the intensity of the pump and $w_{\mu}$ is its half-width. In this Letter we use the pump well above the threshold $\mu=3$ with the half-width $w_{\mu}=0.25$ and $\mu=3$.

We consider a spatially uniform monochromatic coherent pump $P_{ch}=p_0 \exp(i\delta_0 t)$ where $p_0$ is the amplitude and $\delta_0$ is the frequency of the coherent pump, the frequency of the coherent pump is defined as the detuning of the pump frequency from the cavity cut-off frequency in the center of the potential (in $x=0$ ). Such a pump produces coherent polaritons with frequency $\delta_0$ and density much lower compared to the density of the condensate forming because of the incoherent pump. 

\begin{figure}[ht]
\centering
\includegraphics[width=\linewidth]{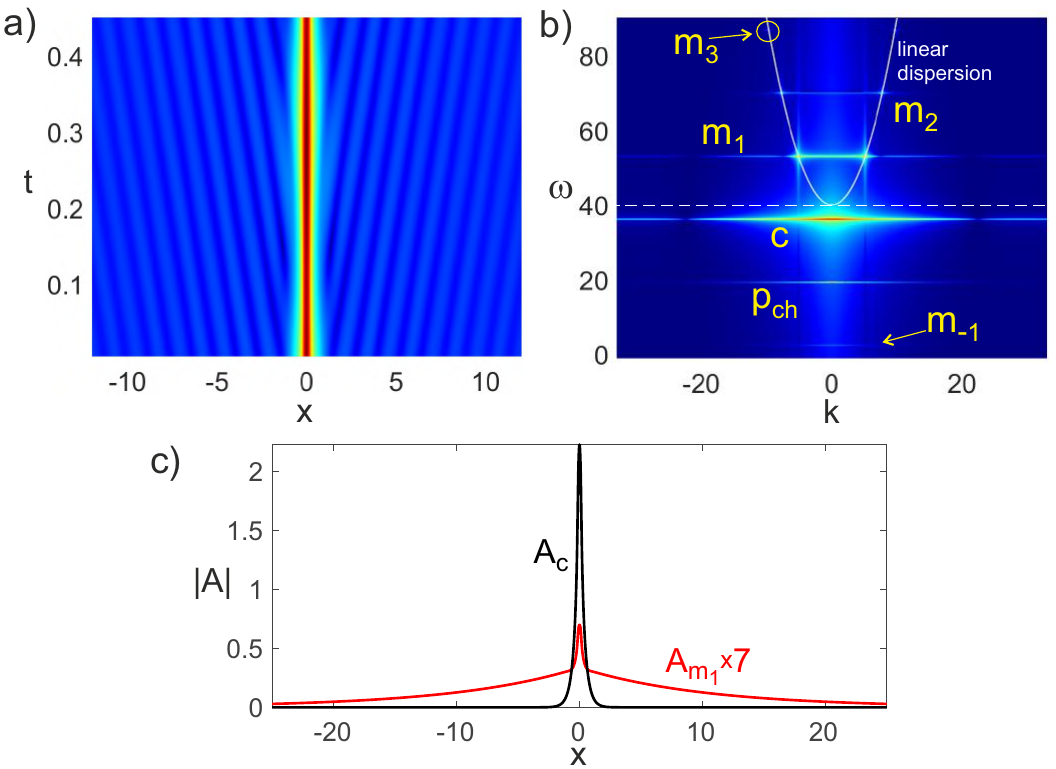}
\caption{ The temporal evolution of the polariton field is shown in panel (a) for the condensates trapped in super-Gaussian potential. The depth of the potential is  $V_0=40$ and its half-width is $W=0.15$, the amplitude and the frequency of the coherent pump are  $p_0=3$ and $\delta=19.3$. The frequencies are defined as the detunings from the cavity cut-off frequency in the center of the potential well.
The polariton condensates are supported by the incoherent pump with the intensity $\mu=3$ and half-width $w_{\mu}=0.25$, see text for the details. The spatio-temporal spectrum of the field is shown in panel (b). The pattern corresponding to the localized potential is marked by yellow $c$, the spectral pattern corresponding to the direct response to the coherent drive is marked by $p_{ch}$. The sidebands appearing due to four wave mixing of  the condensate with the coherent drive are marked as $m_1$-$m_3$ (up-converted) and $m_{-1}$ (down-converted). The wite curve shows the dispersion of the polaritons away from the potential well. The spatial distributions of the filtered fields corresponding to the condensate (black curve) and $m_1$ side-band (red curve) are shown in panel (c).
}
\label{fig1}
\end{figure}

The results of numerical simulations are presented in Fig.~\ref{fig1}. The temporal evolution of the polariton field amplitude is shown in panel (a). It is seen that withing the trap the amplitude of the polariton field is high. 

It has been checked that without the coherent drive $p_0=0$ the polariton field is localized in the trap with very short evanescent tails. If the coherent pump is switched on, the  condensate is still mostly localized within the trapping potential, see Fig.~\ref{fig1}(a) showing the temporal evolution of the polariton field distribution in the presence of the coherent pump with the frequency $\delta_0=19.3$ smaller than the depth of the potential. In the spectral domain shown in Fig.~\ref{fig1}(b) there is a bright spot marked $c$ that can be associated with the trapped part of the condensate. The maximum of the spectral pattern is situated in the vicinity of $k=0$ as it should be for the ground state.

However, as it is seen in Fig.~\ref{fig1}(a) long oscillating tails with fringes appears on the left and right of the condensate in the presence of coherent pump. The tails are the interference of polariton field produced by the coherent drive and the generated side bands at the frequencies $\delta_c +n(\delta_c-\delta_0)$, where $\delta_c$ is the condensate frequency and $n$ is an integer.

In the spectral domain the pattern associated with the hoherent pump is marked as  $p_{ch}$, see panel (b). The down-converted frequencies $n<-1$ are always localized and non-resonant. The frequency of the condensate in our simulations is $\delta_c \approx 36.2$ and thus we can expect the first down-converted spectral line at $\delta \approx 2.4$. Indeed, in  Fig.~\ref{fig1}(b) one can see the fade spectral spot marked as $m_{-1}$ at this frequency.

Contrary to down-converted harmonics, the up-converted frequencies can have the frequency belonging to the continuum spectrum and thus these waves will be emitted and propagate away from the condensate. For our parameter we can expect the up-converted harmonics with the frequencies $\delta_{m1}=53.1$, $\delta_{m_2}=70$, $\delta_{m3}=86.9$. All these harmonics are seen in Fig.~\ref{fig1}(b) at the predicted frequencies, the corresponding patterns are marked as $m_{1}-m_{3}$.

For our parameters all up-converted harmonics have the frequencies belonging to the continuum. That is why the corresponding spectral patterns have maxima lying on the dispersion of the linear waves away from the potential  well. The dispersion of the linear waves is shown by the white parabola and one can see that all up-converted frequencies lies on this curve. The most intensive up-converted harmonic is $m_1$ and the interference fringes of this harmonic with the direct response to the coherent pump is well seen in panel (a) on the left and on the right of the potential well. 

The field of the polaritons having different frequencies can easily be extracted from the numerical simulation data. To do this we take the field in Fourier representation and apply a mask which removes all harmonics outside the specified frequency range. Then we do the inverse Fourier trasform and obtain the coordinate representation of the field. 

To extract the condensate field we set the filtering frequency range $[32,39]$. It is checked that the extracted field is well localized, does not show any temporal dynamics and perfectly coincide with the field of the condensate. The spatial distribution of the filtered field of the condensate is shown in panel (a) of Fig.~\ref{fig1}.

The extracted field of the first up-converted harmonic (the filtering frequency range is $[51,55]$ ) is shown in panel (c) by the red curve. It is seen that the field is decaying to the left and to the right, but the localization range is much longer compared to that of the condensate. The decay rate of the harmonic is defined by the product of the group velocity of the wave and the life time of polaritons. In our dimensionless units this propagation length is $\Lambda=v_g=\sqrt{2(\delta_m-V_0)}$ where $\delta_m$ is the frequency of the wave. The higher harmonics have higher frequencies and thus they propagate for longer distances.

Now let us study the interaction between the condensates mediated by the propagating up-converted harmonics. In numerical simulations we use two-well super-Gaussian potential, the identical wells are separated by the distance $2 x_s$: 
$$V=V_0 \left( 1-\exp\left( -\frac{(x-x_s)^{12}}{w^{12}} \right)-\exp\left( -\frac{(x+x_s)^{12}}{w^{12}} \right)\right). $$
The incoherent pump is $P=\mu \left(\xi \exp\left( - \frac{(x-x_s))^{12}}{w_{\mu}^{12}} \right)+\exp\left( - \frac{(x+x_s))^{12}}{w_{\mu}^{12}} \right)  \right)$ where the coefficient $\xi$ accounts for the right condensate can be pumped stronger or weaker then the left one. The coherent pump and the parameters  $V_0$,  $w$, $\mu$, $w_{\mu}$ are the same as in the case of one-well potential discussed above.

\begin{figure}[ht]
\centering
\includegraphics[width=\linewidth]{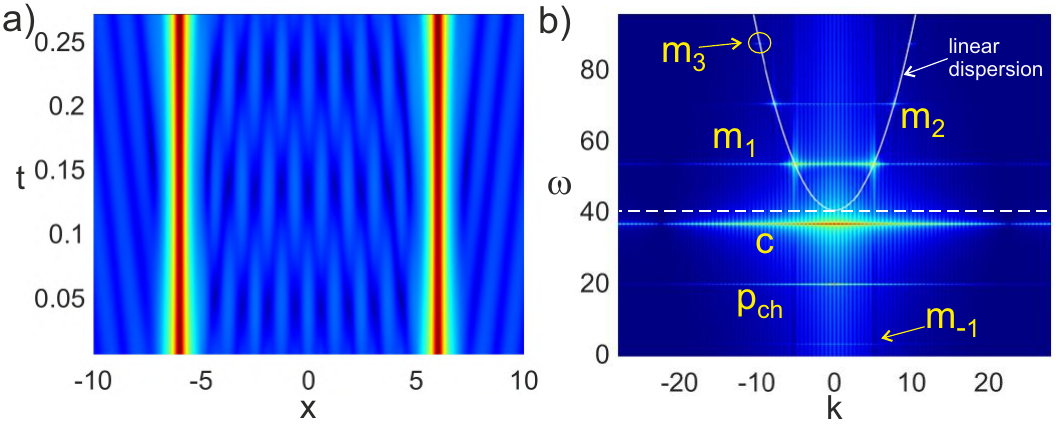}
\caption{ The spatial evolution of the field and the corresponding spatio-temporal spectrum for two condensates forming the two wells super-Gaussian potential.
The parameters are the same as in Fig.~\ref{fig1}, see text for details. The spectral patterns marking is also the same as in Fig.~\ref{fig1}. }
\label{fig2}
\end{figure}

The temporal field evolution in the stationary regime and the corresponding spatio-temporal spectrum are shown in  Fig.~\ref{fig2}(a) and (b) correspondingly. The distance between the condensates is $2 x_s=12$ which is much greater than the localization length of the condensate but less compared to the propagation distance of the up-converted harmonics. We also pump the right condensate $0.5$ $\%$ stronger than the left one, $\xi=1.005$. 

It is seen that complex interference fringes appears between the condensates. They are produced by the counter-propagating up-converted harmonics and the coherent pump. The only noticeable feature in the spectral domain is that all spectral patterns are now modulated along $k$ with the period $\frac{\pi}{x_s}$, this is so because the spectrum is now produced by nearly the same condensates separated by the distance $2x_s$.

Because of a slight difference of the pumps the right and the left condensates have slightly different densities and therefor their free running frequencies do not coincide.  The frequencies of the up-converted harmonics have to be slightly different too. Therefor the intensity of the filtered field of the up-converted harmonic must vary in time with the period equal to the $\frac{\pi}{\Delta \delta}$ where $\Delta \delta$ is the frequency difference between the right and the left condensates. This is exactly what is observed for the coherent pumps of very low intensities. However, for the coherent pump intesity used in the simulations this is not the case.

If the amplitude of the coherent pump exceeds some threshold value then the field becomes time independent which signals that the frequencies of the condensates are now exactly the same. So what is observed is the synchronization of the condensates. Let us remind that the condensate fields are well localized and the direct interaction between the condensates is absolutely negligible. The frequencies of the condensates and the coherent pump is different as it is seen in Fig.~\ref{fig2}(b), therefore the effect cannot be explained by the synchronization of both condensates to the pump. 

\begin{figure}[ht]
\centering
\includegraphics[width=\linewidth]{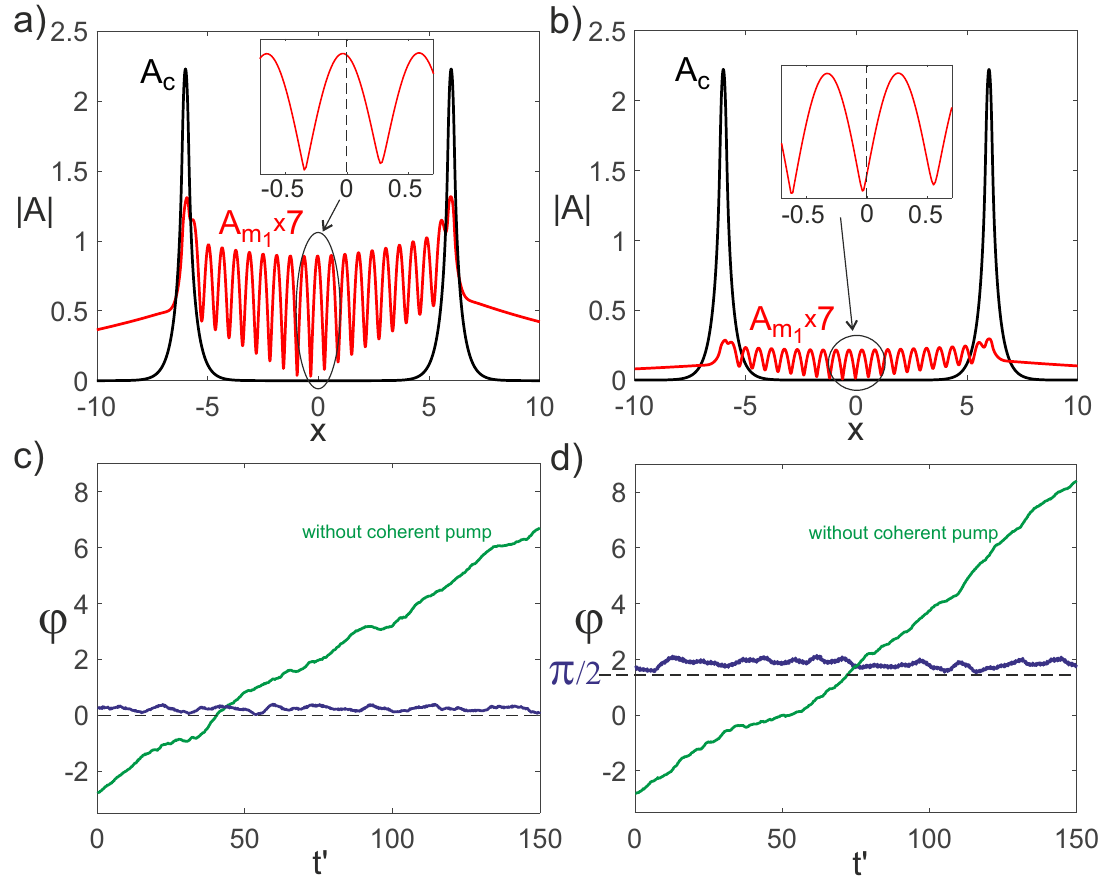}
\caption{ The filtered filed distributions are shown in panels (a) for the system parameters as in Fig.~\ref{fig2}. The black and red curves correspond to the condensate and the first up-converted side band (marked as $c$ and $m_1$ in Fig.~\ref{fig2}). The filed for $m_1$ in the middle between the condensates is shown in the inset where the vertical dashed line marks $x=0$. Panel (b) shows the same as panel (a) but for the coherent pump frequency $\delta=17.9$. The evolution of the mutual phase $\varphi$ of the condensates are shown in panels (c) ( $\delta=19.3$) and (d) ($\delta=17.9$) by the blue curves for the case where the direct drive contains the coherent pump contains and weak  noise. The equilibrium values of the mutual phases are shown by the horizontal black dashed lines. The typical evolution of the mutual phases in the absence of the coherent drive are shown by the green curves. }
\label{fig3}
\end{figure}

The stationary state depends on the frequency of the coherent pump and varying the pump frequency it is possible to achieve different regimes of synchronization, compare the fields shown in Fig.~\ref{fig3}(a) and (b) for the coherent pump frequencies $\delta=19.3$ and $\delta=17.9$. It is worth noting that in the middle between the condensates the field is nearly at maximum for $\delta=19.3$ and nearly at the minimum for $\delta=17.9$. If the condensates are pumped equally then it will be exactly maximum or minimum. This very much resembles the effect observed in for the settings where the condensate have radiating tails. We have studied that as in the synchronization regime depends quasi-periodically on the distance between the condensates. However in the presence of the coherent pump the synchronization also depends quasi-periodically  on the coherent pump frequency.

The synchronization can be conveniently described in terms of the dynamics of the mutual phase between the condensates that can be defined as $\varphi=\arg\left( A(t, x=x_s) A^{*}(t, x=-x_s) \right)$. 
We studied the dynamics of the mutual phase in our system with admixture of weak noise to the coherent pump. 
It is seen in Fig.~\ref{fig3}(c),(d) showing the dynamics of the mutual phases that the in the presence of the coherent pump the mutual phases slightly fluctuate around some values. The green lines show the phase dynamics in the absence of the coherent pump and it is seen that the mutual phase growth in time. Therefore we can conclude that the coherent pump results in the phase locking of the condensates. 
It is interesting to note that the observed synchronization is always bistable. For $\delta=19.3$ the mutual phase is close either to $0$ or to $\pi$ whereas for $\delta=17.9$ the stable equilibrium points are close to $\pm \frac{\pi}{2}$. The positions of the equilibrium points depend on the imbalance of the incoherent pumps acting on the left and right condensates.

Let us explain why the stable synchronization can occur with phases $0$ and $\pi$ (or $\pm \frac{\pi}{2}$, for simplicity we assume that that the condensates are pumped equally). A condensate trapped in a potential well can be seen as a nonlinear oscillator. Let us denote the condensate field in the first and the second trap as $\psi_{1,2}=\rho e^{i\theta_{1,2}}$ where $\rho$ is the real amplitude and $\theta$ is the phase of the condensate. The field of coherent pump we denote as $y=|y|e^{i\theta_c}$, $\theta_c$ is the phase of coherent pump.

Lets consider how the  field emitted by the first condensate affect the second one. The strongest propagating field is the first up-converted harmonics and its field produced by cubic nonlinearity can be expressed as $\psi_1^\cdot y^{*}=\rho^2|y|e^{2i\theta_1-\theta_{c}}$.  When the harmonic arrives to the second condensate it acquires some additional phase $\chi$ and thus the field at the position of the second condensate is $z_1=\rho^2|y|e^{2i\theta_1-\theta_{c}+i\chi}$.

However the field $z_1$ cannot get synchronized with the second condensate directly because of large frequency detuning. But this field can be mixed with the second condensate and the coherent pump which convert the frequency down, the process is $\tilde z_1=z_1\cdot y \cdot\psi_2^{*}$. Thus we get the signal $\tilde z_1=\rho^4|y|^2 e^{2i\theta_1-\theta_{2}+i\chi}$ with the frequency equal to the frequency of the condensate. 

The interaction of this signal $\tilde z=|\tilde z|e^{i\theta_{\tilde z}}$, $\theta_{\tilde z}=2\theta_1-\theta_2+\chi$ to the second condensate can be described by Adler equation for the phase dynamics $\dot \theta_2=u_s \sin (\theta_2-\theta_{\tilde z})+u_c\cos(\theta_2-\theta_{\tilde z})$ where $u_s$ and $u_c$ are real constants that can be expressed through the amplitude $|\tilde z|$ and the parameters of the system. The analogous procedure gives the equation for the phase of the first condensate. 

Finally we arrive to the equation for the mutual phase $\varphi=\theta_2-\theta_1$ in the form 
\begin{eqnarray}
\dot \varphi =2u_s\sin(2\varphi) \label{synchr}
\end{eqnarray}
 The difference from the well known equation for the mutual phase is  the coefficient $2$ standing in front of the argument of $\sin$ function.

 Equation (\ref{synchr}) has four equilibrium points $\varphi=0$, $\varphi=\pi$ and $\varphi=\pm\frac{\pi}{2}$. The equilibrium points $\varphi=0$ and $\varphi=\pi$ are both stable  if $u_s<0$, for $u_s>0$ the stable points  are $\varphi=\pm\frac{\pi}{2}$. The sign of $u_s$ depends on many factors, in particular on the frequency of the coherent pump. So this explains why in the numerical simulations the synchronization with the mutual phase $\pm \frac{\pi}{2}$ was observed. A more comprehensive theory of the synchronization will be published later as a separate paper.

To conclude, we demonstrated that two separated polariton condensates can be synchronized due to the interaction mediated by propagating polariton produced by the four wave mixing of the polariton condensates with the external coherent pump. The  consequent up and down conversions of the polaritons is crucial for the synchronization to occur and thus it is sensitive not only to the distance between the condensates and their frequencies but also to the frequency of the coherent pump. It is interesting to note that the stationary mutual phase depends on the initial conditions. For some coherent pump frequencies the stationary phase can be equal either to $0$ or $\pi$, for the other frequencies the synchronization occurs with either $\varphi=\frac{\pi}{2}$ or $\varphi=-\frac{\pi}{2}$.

The effect is demonstrated for the simplest possible configuration but can easily be generalized for the wide class of the systems. In particular, to increase the synchronization strength the coherent pump can be focused in two spots. The discussed phenomenon can be of interest from the point of view of the control and manipulation of the exciton-polariton condensates, in particular in the lattices of the condensate droplets. 

\bigskip
\bigskip

Acknowledgments

\bigskip

 This work was supported by  Russian Science Foundation,  grant 23-72-00031.
 
\bigskip
\bigskip

% % Bibliography

\end{document}